\documentstyle[aps,preprint,epsf]{revtex}

\def\beq{\begin{equation}}
\def\eeq{\end{equation}}
\def\bea{\begin{eqnarray}}
\def\eea{\end{eqnarray}}
\newcommand{\sla}{\!\!\!\!/ \,}

\begin{document}

\draft

\title{One-loop self energies at finite temperature\footnote{Supported by BMBF,
GSI Darmstadt, and DFG}}
\author{Andr\'e Peshier$^{1,2}$, Klaus Schertler$^3$, and Markus H.
Thoma$^{3,4}$\footnote{Heisenberg fellow}}
\address{$^1$Research Center Rossendorf Inc., PF 510119, 01314 Dresden,
Germany\\
$^2$Institut f\"ur Theoretische Physik, Technische Universit\"at
Dresden, Mommsenstr. 13, 01062 Dresden, Germany\\
$^3$Institut f\"ur Theoretische Physik, Universit\"at Giessen,
35392 Giessen, Germany\\
$^4$European Center for Theoretical Studies in Nuclear Physics and Related
Areas, Villa Tambosi, Strada delle Tabarelle 286, I-38050 Villazzano, Italy}
\maketitle

\begin{abstract}
The complete one-loop self energies (real and imaginary parts) for photons,
gluons, electrons and quarks at finite temperature are calculated numerically
and compared to the results of the hard thermal loop (HTL) approximation used
for the resummation technique of Braaten and Pisarski.
In this way some light is shed on the validity of the weak coupling limit
assumption ($g \ll 1$) or equivalently the high temperature assumption,
on which the HTL approximation is based. Furthermore, the gauge dependence of
the fermion self energy beyond the HTL approximation is considered.
Finally the dispersion relations following from the real part of the self
energies are compared to the HTL results.
\end{abstract}


\section{Introduction}

In relativistic plasmas, such as a quark-gluon plasma (QGP) possibly formed
in relativistic
heavy ion collisions or a QED plasma in a supernova explosion, and in relativistic,
degenerate matter, as e.g. strange quark stars, medium effects are of great importance.
For example, the originally massless or light particles of the system acquire an effective
mass caused by the interactions with the other particles of the system. Also screening
effects due to the presence of (color) charges in the system (Debye screening) prevent
or at least reduce
infrared singularities coming from the long range interactions in QED or QCD.

These effects can be calculated from the self energy of the particles.
Evaluated perturbatively to lowest order at high temperature $T$ or density
(corresponding to a large chemical potential $\mu $) the effective mass and the
Debye mass are of the order $gT$ or $g\mu $, where $g$ is the coupling constant.
These masses can be large compared to the bare masses, which are negligible at
high temperature.
The dispersion relations of the quasiparticles thus deviate
from the ones of  bare particles leading to important modifications of the
equation of state \cite{Peshier,Schertler}. Therefore medium effects are an
essential feature of relativistic systems at finite temperature or density.

So far these medium effects have mainly been studied by considering the self energies
in the high temperature approximation \cite{Klimov,Weldon1,Weldon2}
or, equivalently, in the hard thermal loop (HTL) limit \cite{BP}.
In this approximation the momenta of the internal particles of the self
energy are assumed to be hard, i.e.\ of order of the temperature or chemical potential,
whereas the external momenta are soft, i.e.\ of the order $gT$ ($g\mu $). These momentum
scales can be distinguished only in the weak coupling limit $g\ll 1$, on which the
HTL approximation is based.
The reason for these restrictions is twofold.
First this approximation allows for analytic expressions of the self energies. Secondly,
which is more important, the self energies calculated in this way are gauge independent
guaranteeing gauge independent results for the effective masses, the dispersion relations,
and the Debye screening. As a matter of fact, the HTL approximation builds the basis for a
consistent extension of the naive perturbation theory, which suffers from infrared
singularities and gauge dependent results for observables. Resumming the HTL self energies
(and vertices) leads to effective Green's functions, which have to be used in the case of soft
external momenta (Braaten-Pisarski method \cite{BP}). In this way consistent results for
physical quantities can be found, which are gauge independent and complete to leading order
in the coupling constant. At the same time the infrared behavior is greatly improved due
to the use of resummed propagators, although there might be still infrared divergences
owing to the absence of a static magnetic screening in the HTL self energies of the
gauge bosons. For applications of this powerful method to interesting problems of the
QGP we refer to Ref.\cite{QGP}.

Here we want to compute the self energies and dispersion relations of photons, gluons
and fermions at finite temperature but zero chemical potential within the one-loop
approximation abolishing the HTL assumption and compare the results to the ones of the
HTL approximation.
This will give us a hint for the validity of the weak coupling assumption, on which
the HTL approximation and thus the Braaten-Pisarski method relies.
Up to now there are only explicit expressions for the gluon self energy beyond the
HTL limit, where the next term of the high temperature expansion has been evaluated
\cite{Toimela1}.
For a complete one-loop calculation, which can be done only numerically, there exist
only integral expressions \cite{Kapusta}, results for certain momentum limits
\cite{Kajantie}, or analytic expressions for zero temperature but non-zero chemical
potential \cite{Toimela2}.
Furthermore, it is instructive to have a look at quantities which vanish at leading order
of the one-loop high-$T$ expansion. As examples we consider here the imaginary parts
of the self energies above the light cone and the gauge dependence of the fermion
self energy.

\section{Photon self energy}

First we consider the photon self energy or polarization tensor of QED within the one-loop
approximation.
In the vacuum it reads ($K\equiv (k_0,{\bf k})$, $k\equiv |{\bf k}|$)
\begin{equation}
 \Pi_{\mu\nu}^{QED}(P)
 =
 -i e^2 \int \frac{d^4 K}{(2 \pi)^4} tr[ \gamma_\mu S(-Q) \gamma_\nu S(K) ]
\label{e1}
\end{equation}
where $S$ denotes the electron propagator and $P=K+Q$.
In order to compute $\Pi_{\mu\nu}^{QED}$ at finite temperature we apply the imaginary time
formalism and, for temperatures $T \gg m_e$, neglect the electron mass, i.e.
$S(K) = K\sla \Delta _F(K)$ where $\Delta _F(K)=1/K^2$ with
$k_0=i(2n+1)\pi T$ for integers $n$.
To perform the sum over the Matsubara frequencies $k_0$
it is convenient to use the Saclay representation of the propagators
\cite{Pisarski}:
\begin{equation}
\Delta_F(K) = - \int_0^\beta d\tau e^{k_0 \tau} \Delta_F(\tau,k)
\label{e2}
\end{equation}
with
\begin{equation}
 \Delta_F(\tau,k) = \frac{1}{2k} \left\{ [1-n_F(k)] e^{-k \tau} - n_F(k) e^{k \tau} \right\},
\label{e3}
\end{equation}
where $n_F(k)=1/[\exp (k/T)+1]$ denotes the Fermi distribution.

Defining $\Pi_L^{QED} = \Pi_{00}^{QED}$ and
$\Pi_T^{QED} = \frac12\, (\delta_{ij}-\hat p_i\hat p_j) \Pi_{ij}^{QED}$,
($\hat p \equiv {\bf p}/p$), the longitudinal and transverse projections, respectively, of
the self energy read
\bea
 \Pi_L^{QED}(P)
 &=&
 - 4 e^2 \int \frac{d^3 k}{(2 \pi)^3} \, T \sum_{k_0} (k_0 q_0 + {\bf k} \cdot {\bf q})
         \Delta_F(Q)  \Delta_F(K)
 \nonumber \\
 \Pi_T^{QED}(P)
 &=&
 - 4 e^2 \int \frac{d^3 k}{(2 \pi)^3} \, T \sum_{k_0}
         \left( k_0 q_0 - ({\bf\hat p}\cdot{\bf k})({\bf\hat p}\cdot{\bf q}) \right)
          \Delta_F(Q)  \Delta_F(K) \, .
 \label{e4}
\eea
In the HTL approximation, i.e.\ assuming $p_0, p \ll T$, analytic expressions can be obtained
\bea
 \Pi_L^{\rm HTL}(P)
 &=&
  -3 m_\gamma^2 \left[ 1-\frac{p_0}{2 p}\ln\frac{p_0+p}{p_0-p} \right],
 \nonumber \\
 \Pi_T^{\rm HTL}(P)
 &=&
  \frac32\, m_\gamma^2\, \frac{p_0^2}{p^2}
  \left[ 1- \left( 1-\frac{p^2}{p_0^2} \right) \frac{p_0}{2 p}\ln\frac{p_0+p}{p_0-p} \right],
 \label{e5}
\eea
where $m_\gamma=eT/3$ denotes the effective, thermal photon mass. The derivation of (\ref{e5})
can be found, e.g., in the appendix of ref.\cite{QGP}.

Now we want to evaluate the complete one-loop self energies starting from (\ref{e4}), which we will write as
\bea
 \Pi_L^{QED}(P)
 &=&
  - 4 e^2 \int \frac{d^3 k}{(2 \pi)^3} \,
   \left[ F_{F,F}^{(1,1)} + {\bf k} \cdot {\bf q} \, F_{F,F}^{(0)} \right] ,
 \nonumber \\
 \Pi_T^{QED}(P)
 &=&
  - 4 e^2 \int \frac{d^3 k}{(2 \pi)^3} \,
  \left[
    F_{F,F}^{(1,1)} - ({\bf\hat p}\cdot{\bf k})({\bf\hat p}\cdot{\bf q})\, F_{F,F}^{(0)}
  \right] ,
\label{e6}
\eea
where
\begin{eqnarray}
 F_{F,F}^{(1,1)}
 &=&
 T \sum_{k_0} k_0 q_0 \Delta_F(K) \Delta_F(Q)
 \nonumber \\
 &=&
 \frac{1}{4} \left\{ [1 - n_F(q) - n_F(k)] \left( \frac{1}{p_0+k+q} - \frac{1}{p_0-k-q}
 \right) \right.
 \nonumber \\
 & &
 \quad
 +\left. [n_F(q) - n_F(k)] \left( \frac{1}{p_0+k-q} - \frac{1}{p_0-k+q} \right) \right\}
\label{e7}
\end{eqnarray}
and
\begin{eqnarray}
 F_{F,F}^{(0)}
 &=&
 T \sum_{k_0} \Delta_F(K) \Delta_F(Q)
 \nonumber \\
 &=&
 \frac{1}{4 k q} \left\{ [1 - n_F(q) - n_F(k)]
 \left( \frac{1}{p_0+k+q} - \frac{1}{p_0-k-q} \right) \right.
 \nonumber \\
 & &
 \qquad
   -\left. [ n_F(q) - n_F(k)] \left( \frac{1}{p_0+k-q} - \frac{1}{p_0-k+q} \right) \right\}.
\label{e8}
\end{eqnarray}

Next we neglect the vacuum parts, i.e.\ the terms which do not contain any distribution
functions in (\ref{e7}) and (\ref{e8}), as we are interested only in the temperature
dependent medium effects. The angle integration over $\eta \equiv \hat {\bf p} \cdot \hat
{\bf k}$ can be performed analytically. (For this purpose we remove the $\eta $-dependence
in $n_F(q)$ by shifting ${\bf k}\rightarrow {\bf p}-{\bf k}$ in the corresponding terms.)
The remaining integral over $k$ has to be done numerically.
The final results for the real part of the self energy are shown in Fig.\ 1 and Fig.\ 2,
where they are also compared to the HTL results (\ref{e5}).
For the longitudinal component, we observe a surprisingly good agreement between the two
approaches even when the HTL assumption $p_0,p \ll T$ is not fulfilled.
The transverse component, however, is more sensitive to the violation of the HTL assumption.
For small $p$ and $p_0$ the one-loop result approaches the HTL limit as expected.
Obviously, near the light cone the results of both approaches coincide even outside the range
of validity of the high-$T$ assumption. Indeed, the behavior of, e.g., the longitudinal self
energy, for $p_0 \sim p$ and {\em all} values of $p$ is readily found to be
$\Pi_L^{QED} \sim 3 m_\gamma^2\, p_0/(2p) \ln[(p_0+p)/(p_0-p)]$, which is just the HTL result
(\ref{e5}) extrapolated to arbitrary, nearly light-like momenta.

Next we consider the dispersion relations, which are given by the poles of the
effective photon propagator following from resumming the self energy. For the
longitudinal mode (plasmon), which is absent in the vacuum, and the
transverse mode, respectively, they are determined by
\bea
 {D_L^{*\, QED}(P)}^{-1}
 &=&
 p^2-\Pi_L^{QED}(P)=0
 \nonumber \\
 {D_T^{*\, QED}(P)}^{-1}
 &=&
 P^2-\Pi_T^{QED}(P)=0 \, .
\label{e9}
\eea
As in the vicinity of the light cone the deviations of both approaches are small,
the same is expected for the dispersion relations for not too large couplings.
Indeed, even in the case of extrapolating the coupling constant to $e=3$, the
difference is less than 5\% (Fig.\ 3).
We note that in a good approximation the one-loop dispersion relations follow
>from the HTL approximations by a suitable rescaling of $m_\gamma$.
As expected, both branches meet for $p=0$ in either approximation.

\section{Gluon self energy}

Here we compute the complete one-loop gluon self energy which, in contrast
to the abelian boson polarization, is gauge dependent.
In Feynman gauge, the pure gauge contribution of the gluon loop, the gluon
tadpole and the ghost loop diagram is \cite{Weldon1,BP}
\begin{equation}
 \Pi_{\mu\nu}^{QCD}(P)
 =
 -3 g^2 \int \frac{d^3 k}{(2 \pi)^3} \, T \sum_{k_0}
 \left[
   P^2 g_{\mu\nu}-P_\mu P_\nu + 2(KQ g_{\mu\nu}-K_\mu Q_\nu-K_\nu Q_\mu)
 \right]
 \Delta_B(K)  \Delta_B(Q) \, ,
 \label{e30}
\end{equation}
where $\Delta _B(K)=1/K^2$ with $k_0=i 2n\pi T$.
The Saclay representation of the bosonic propagator reads
\begin{equation}
 \Delta_B(K) = - \int_0^\beta d\tau e^{k_0 \tau} \Delta_B(\tau,k) \, ,
 \label{e12}
\end{equation}
with
\begin{equation}
 \Delta_B(\tau,k)
 =
 \frac{1}{2k} \left\{ [1+n_B(k)] e^{-k \tau} + n_B(k) e^{k \tau} \right\}
 \label{e13}
\end{equation}
and the Bose distribution $n_B(k) = 1/[\exp(k/T)-1]$.
As for the photon, we consider the longitudinal and transverse components
of $\Pi_{\mu\nu}^{QCD}$. Expressed by the functions $F_{B,B}^{(1,1)}$,
$F_{B,B}^{(0)}$ defined in analogy to the fermionic functions in (\ref{e7},\ref{e8})
with $-n_F$ replaced by $n_B$, they read
\bea
 \Pi_L^{QCD}
 &=&
 3 g^2 \int \frac{d^3 k}{(2 \pi)^3} \,
 \left[
   2 F_{B,B}^{(1,1)} + (2{\bf k}\cdot{\bf q}+p^2) \, F_{B,B}^{(0)}
 \right] ,
 \nonumber \\
 \Pi_T^{QCD}
 &=&
 3 g^2 \int \frac{d^3 k}{(2 \pi)^3} \,
 \left[
   2 F_{B,B}^{(1,1)}
 + (p_0^2-p^2
    -2({\bf\hat p}\cdot{\bf k})({\bf\hat p}\cdot{\bf q})) \, F_{B,B}^{(0)}
 \right] .
 \label{e31}
\eea
Under the HTL assumption, terms $\propto p_0^2,p^2$ drop out
and the integrals over the bosonic functions yield up to a factor of two
the same expressions as for the fermionic case, so
the self energies obviously approximate to just the abelian expressions
(\ref{e5}) with $m_\gamma$ now replaced by the thermal gluon mass
$m_g=gT/\sqrt3$.

In comparison to the abelian case for the same values of $p_0, p$, the difference
to the HTL result is evidently larger (Fig.\ 4 and Fig.\ 5). Again we find that
for soft momenta as well as for $p_0 \sim p$ the HTL limit is approached and
that the transverse part is more sensitive to the overstraining of the HTL assumption.

In general, beyond the leading contribution of the high-$T$ expansion, the (one-loop)
self energies have a non-vanishing imaginary part above the light cone, for $p_0>p$.
For the gluon self energies it is shown in Figs.\ 6 and 7.
While the general requirement $\Im(\Pi)<0$, corresponding to a positive damping rate,
is fulfilled for the one-loop photon self energy, it is violated by, e.g., the
transverse gluon self energy (in Feynman gauge).\footnote{
 Historically, this was one of the main triggers for the development of the Braaten-Pisarski
 method.}
This inconsistency indicates that extending the high-$T$ expansion beyond leading order
requires higher-loop contributions to obtain a consistent propagator.
The positive imaginary parts above the light cone thus provide a measure of the magnitude
of these sub-leading terms which may be of a similar order as the imaginary HTL contributions
for $p_0<p$.

Although self energies containing gluon propagators depend on the choice of the gauge,
the dispersion relations are, in principle, observable and thus gauge independent.
While the gauge dependent terms of the gluon self energy are (somewhat surprisingly)
found to be of sub-leading order in the high-$T$ expansion and the HTL-dispersion
relations are thus gauge invariant, this is not the case in the one-loop approximation.
Postponing the question of the gauge effects to the next section, here in Feynman gauge
the influence of the sub-leading terms neglected in the HTL-limit is considered.
For the dispersion relations of the longitudinal and transverse gluons we find a
decreasing gluon energy $\omega(p)$ with increasing $g$.
This qualitatively resembles to the next-to-leading order behavior of the gluon mass
\cite{Schulz}.
The deviations from the HTL result are larger than in the case of photons for the
same extrapolated coupling strength (Fig.\ 8).

\section{Fermion self energy}

Here we consider the electron or quark self energy within the one-loop approximation.
At zero temperature it is given by
\begin{equation}
\Sigma(P) = i g^2 C_R \int \frac{d^4 K}{(2 \pi)^4} D_{\mu \nu}(Q) \gamma^\mu S(K) \gamma^\nu,
\label{e10}
\end{equation}
where the Casimir invariant of the fundamental representation is $C_R=4/3$ for quarks
(for electrons, $g^2 C_R \rightarrow e^2$).
We will first choose the Feynman gauge for the gauge boson propagator,
\begin{equation}
 D_{\mu \nu}(Q) = -g_{\mu \nu} \frac{1}{Q^2} = -g_{\mu \nu} \Delta_B(Q) \, .
\label{e11}
\end{equation}
Combining (\ref{e10}) and (\ref{e11}), the finite temperature self energy is given by
\begin{equation}
\Sigma(P) = -2 g^2 C_R \int \frac{d^3 k}{(2 \pi)^3} \enspace T \sum_{k_0}  K\sla  \Delta_F(K)
\Delta_B(Q).
\label{e14}
\end{equation}

For evaluating this expression we utilize a general representation of the fermion self energy
in the rest frame of the heat bath \cite{Weldon2}:
\begin{equation}
\Sigma(P) = -a P\sla -b \gamma_0
\label{e15}
\end{equation}
with the scalar functions
\begin{eqnarray}
a(P) & = & \frac{1}{4 p^2} \left(tr[P\sla \Sigma] - p_0 tr[\gamma_0 \Sigma] \right) \nonumber \\
b(P) & = & \frac{1}{4 p^2} \left( P^2 tr[\gamma_0 \Sigma] - p_0 tr[P\sla \Sigma]  \right) .
\label{e16}
\end{eqnarray}

Evaluating the traces
\begin{eqnarray}
 T_1(P) & \equiv & tr[\gamma_0 \Sigma(P)] \, ,
 \nonumber \\
 T_2(P) & \equiv & tr[P\sla \Sigma(P)] \, ,
\label{e17}
\end{eqnarray}
we are able to determine the functions $a$ and $b$. Performing the sum over $k_0$ we get
\begin{equation}
T_1(P) = - 8 g^2 C_R \int \frac{d^3 k}{(2 \pi)^3} \enspace F_{F,B}^{(1)} \, ,
\label{e18}
\end{equation}
where
\begin{eqnarray}
 F_{F,B}^{(1)}
 &=&
  -\frac{1}{4 q} \left\{ [1+n_B(q) - n_F(k)]
 \left( \frac{1}{p_0+k+q} + \frac{1}{p_0-k-q} \right) \right.
 \nonumber \\
 & & \qquad
  + \left. [n_B(q) + n_F(k)] \left( \frac{1}{p_0+k-q} + \frac{1}{p_0-k+q} \right)
 \right\}.
\label{e19}
\end{eqnarray}
Again we neglect the vacuum contribution. After integrating analytically over the angle
and numerically over $k$ we end up with the final result for $T_1$.
In Fig.\ 9 the real part of $T_1$ is compared to the HTL result
\begin{equation}
 T_1^{\rm HTL}(P) = 2 m_f^2\, \frac{1}{p}\, \ln\frac{p_0+p}{p_0-p} \, ,
\label{e20}
\end{equation}
which contains the effective, thermal fermion mass
\begin{equation}
 m_f^2 = C_R\, \frac{g^2 T^2}{8} \, .
\label{e21}
\end{equation}
As for the boson self energies we observe the one-loop results approaching the HTL predictions
for $p_0,p \ll T$ and $p_0 \sim p$ (Fig.\ 9).

Similar results can be found for $T_2$, which follows from
\begin{eqnarray}
T_2(P) & = & 8 g^2 C_R \int \frac{d^3 k}{(2 \pi)^3} \, \left[ {\bf p} \cdot {\bf k} \,
F_{F,B}^{(0)} - p_0 \, F_{F,B}^{(1)} \right],
\label{e22}
\end{eqnarray}
where
\begin{eqnarray}
 F_{F,B}^{(0)}
 &=&
 \frac{1}{4 k q} \left\{ [1+n_B(q) - n_F(k)]
 \left( \frac{1}{p_0+k+q} - \frac{1}{p_0-k-q} \right) \right.
 \nonumber \\
 & & \qquad
  + \left. [n_B(q) + n_F(k)] \left( \frac{1}{p_0+k-q} - \frac{1}{p_0-k+q} \right)
 \right\}.
\label{e23}
\end{eqnarray}

The comparison of the real part of $T_2$ with the HTL result
\begin{equation}
T_2^{\rm HTL}(P)=4 m_f^2
\label{e24}
\end{equation}
is shown in Fig.\ 10.

In contrast to the HTL approximation, both one-loop fermion self energy functions
develop an imaginary part. As shown in Fig.\ 11 and Fig.\ 12, these may be
as large as the real parts.
We note that the relevant combinations $\pm T_2(P)-(\pm p_0-p)T_1(P)$ (see below)
have non-positive imaginary parts.

Next we consider the dispersion relations following from the resummed fermion propagator.
It is convenient to decompose this propagator according to its helicity eigenstates \cite{BPY}:
\begin{equation}
S^*(P) = \frac{\gamma_0-\hat p \cdot \gamma}{2 D_+(P)} + \frac{\gamma_0+\hat p \cdot \gamma}
{2 D_-(P)}
\label{e25}
\end{equation}
        with
\begin{equation}
D_\pm(P) \equiv -p_0 \pm p + \frac{1}{4 p} \left( \pm T_2(P) - (\pm p_0 -p) T_1(P) \right).
\label{e26}
\end{equation}
The dispersion relations following from the roots of (\ref{e26}) are depicted in Fig.\ 13.
The lower branch (plasmino) corresponding to $D_-(P)$ and displaying a minimum at finite
momentum, which has a negative helicity to chirality ratio \cite{BPY}, is absent in the vacuum.
In Fig.\ 13 the complete one-loop dispersion relations (in Feynman gauge) are compared
to the HTL ones.
The differences increase with increasing coupling constant. However, even for $g=3$ the
maximum deviation is of the order of 10\%.

Finally we investigate the gauge dependence of the one-loop fermion self energy. For this
purpose we adopt the gauge boson propagator in a general covariant gauge
\begin{equation}
D_{\mu \nu}(K) = -g_{\mu \nu} \frac{1}{K^2} + \xi K_\mu K_\nu \frac{1}{[K^2]^2}
\label{e27}
\end{equation}
with the gauge fixing parameter $\xi$ (see e.g. \cite{Itzykson}). For treating
the double pole in the second term we introduce a regularization parameter $\lambda $
\cite{Baier} and write
\begin{equation}
\frac{1}{[K^2]^2} = \left. \frac{\partial}{\partial \lambda} \frac{1}{K^2-\lambda}
\right|_{\lambda=0} = \left. \frac{\partial}{\partial \lambda} \Delta_B(K_\lambda)
\right|_{\lambda=0}.
\label{e28}
\end{equation}
Here $K_\lambda$ indicates the four momentum $(k_0, \vec k_\lambda)$ with
$|\vec k_\lambda| = k_\lambda \equiv \sqrt{k^2+\lambda}$. Substituting the gauge dependent
part of the boson propagator (\ref{e27}) into the self energy (\ref{e14}) we obtain
\begin{equation}
\delta\Sigma(P)=-g^2 \xi C_R \left. \frac{\partial}{\partial \lambda}
\int \frac{d^3 k}{(2 \pi)^3} \enspace T \sum_{k_0} K\sla Q\sla K\sla \Delta_B(K_\lambda)
\Delta_F(Q)\right|_{\lambda=0}.
\label{e29}
\end{equation}
After some manipulations similar to the one for computing $T_1$ and $T_2$ we
find the gauge dependent parts $\delta T_1$ and $\delta T_2$ of the complete
one-loop expressions for $T_1$ and $T_2$ as a function of $p_0$ and $p$
(Fig.\ 14 and Fig.\ 15).
The fact that by Ward-identities and power counting arguments \cite{BP}
they are of the same order as typical corrections
of HTL by sub-leading terms is numerically found to hold also for hard momenta
$p_0,p \sim T$.
As expected, the gauge dependent contributions vanish for small $p_0$ and $p$,
where the one-loop self energy approaches the HTL result.

\section{Conclusions}

In this work we have calculated the complete one-loop self energies for
photons, gluons, and fermions (electrons or quarks) and compared them to
their HTL limits. In the case of the gauge independent one-loop photon
self energy we found a good agreement between these two approaches.
Especially for the dispersion relations both approximations agree to a
surprisingly high accuracy.
As far as the real parts of the gluon and fermion self energies are
concerned, the difference between the both approaches is more distinct.
For momenta near the light cone, however, both approaches coincide even
for hard momenta violating the HTL assumption $p_0,p \ll T$.
This means that for the HTL approximation being valid we need not to
require that both $p_0$ and $p$ are soft, but assuming a soft $P^2=p_0^2-p^2$
is sufficient.
This also justifies the use of the HTL resummed quark propagator for soft
four-momenta in the calculation of the production rate of hard photons from
the QGP \cite{KLS}.
For the imaginary parts there are somewhat larger differences between
the complete one-loop and the HTL calculation. In particular the one-loop
fermion self energy shows a significant imaginary part above the light cone
which may be of the same order as the real part. (The same is expected for
the (complete) gluon self energy beyond leading HTL order.)
Finally we have studied the gauge dependence of the one-loop fermion self
energy in the class of covariant gauges.

The results found above indicate that the HTL approximation, based on
the weak coupling limit assumption $g\ll 1$ or
equivalently on the high temperature assumption $T\gg p_0,p$, is qualitatively
correct for a large momentum regime even for large values of $g>1$ or
temperatures of the order $p$.
This gives us some confidence in the usefulness of the effective perturbation
theory (Braaten-Pisarski method), which is based on the HTL resummation,
for realistic values of the coupling constant $g=1.5$ -- 2.5 at temperatures $T=200$ --
300 MeV that are accessible in relativistic heavy ion collisions.

Furthermore, the use of HTL effective masses for describing the equation of state of a gluon
gas \cite{Peshier} or strange quark matter \cite{Schertler} appears to be justified from
the point of view that the deviation of the HTL masses from the complete one-loop
effective masses can be neglected in those models.

\begin{figure}
 \epsfxsize=10cm
 \epsffile{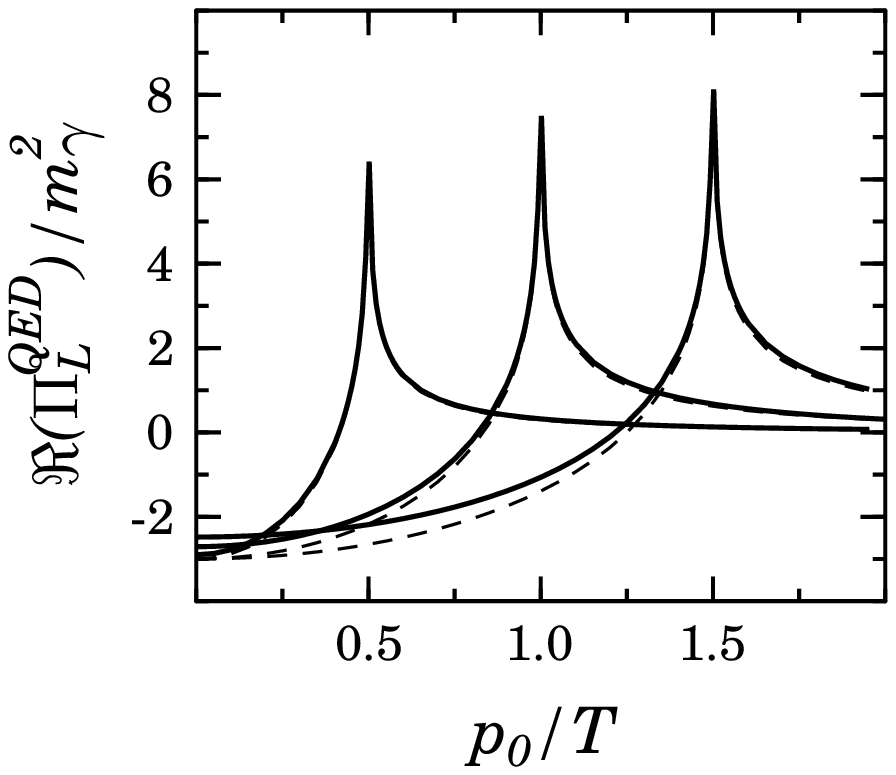}
 \caption{Real part of the longitudinal photon self energy within the complete one-loop
 approximation (solid line) and the HTL limit (dashed line) for momenta $p/T=0.5, 1, 1.5$.}
\end{figure}

\begin{figure}
 \epsfxsize=10cm
 \epsffile{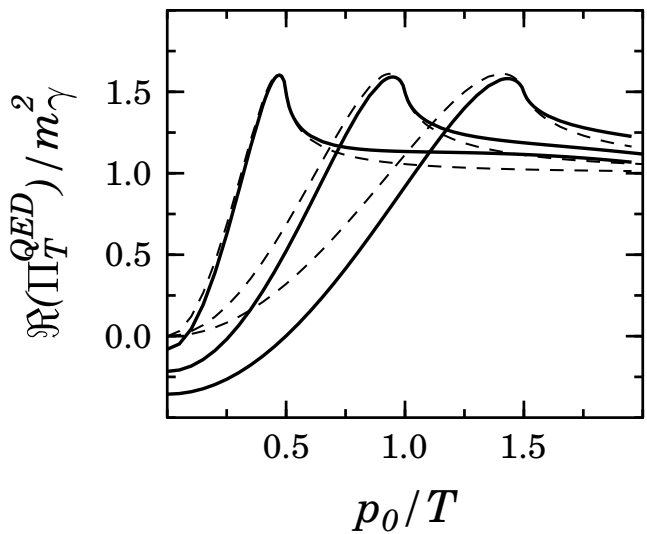}
 \caption{Real part of the transverse photon self energy within the complete one-loop
 approximation (solid line) and the HTL limit (dashed line) for $p/T=0.5, 1, 1.5$.}
\end{figure}

\begin{figure}
 \epsfxsize=10cm
 \epsffile{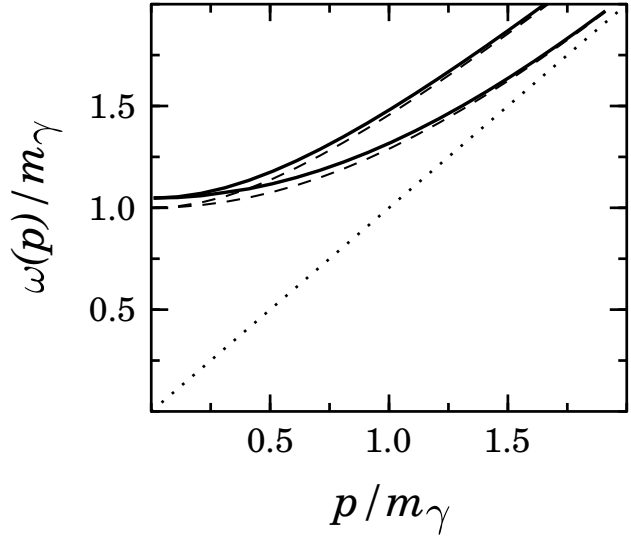}
\caption{Dispersion relation of longitudinal and transverse photons
 (lower and upper branches, respectively)
 within the complete
 one-loop approximation for $e=3$ (solid line) and the HTL limit (dashed line).}
\end{figure}

\begin{figure}
 \epsfxsize=10cm
 \epsffile{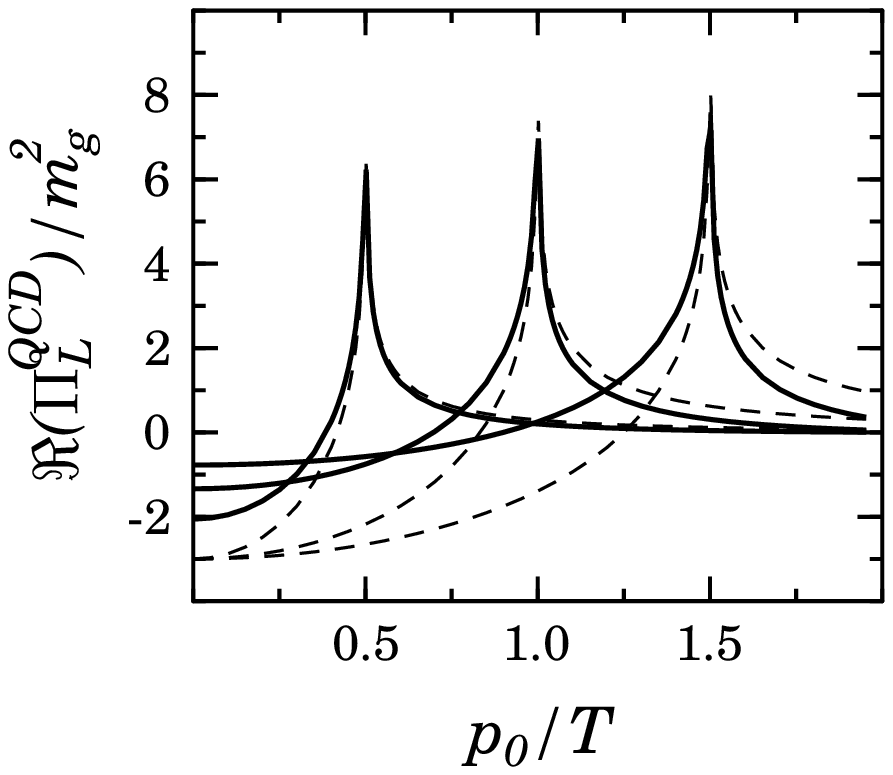}
 \caption{Real part of the longitudinal gluon self energy within the complete one-loop
 approximation in Feynman gauge (solid line) and the HTL limit (dashed line) for
 $p/T=0.5, 1, 1.5$.}
\end{figure}

\begin{figure}
 \epsfxsize=10cm
 \epsffile{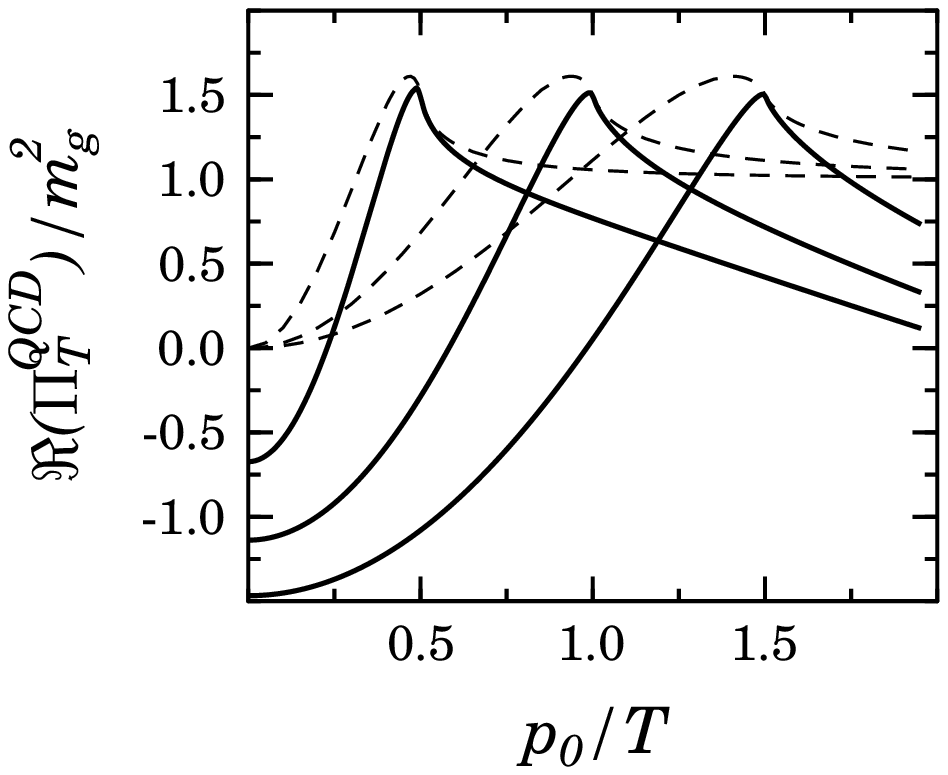}
 \caption{Real part of the transverse gluon self energy within the complete one-loop
 approximation in Feynman gauge (solid line) and the HTL limit (dashed line) for
 $p/T=0.5, 1, 1.5$.}
\end{figure}

\begin{figure}
 \epsfxsize=10cm
 \epsffile{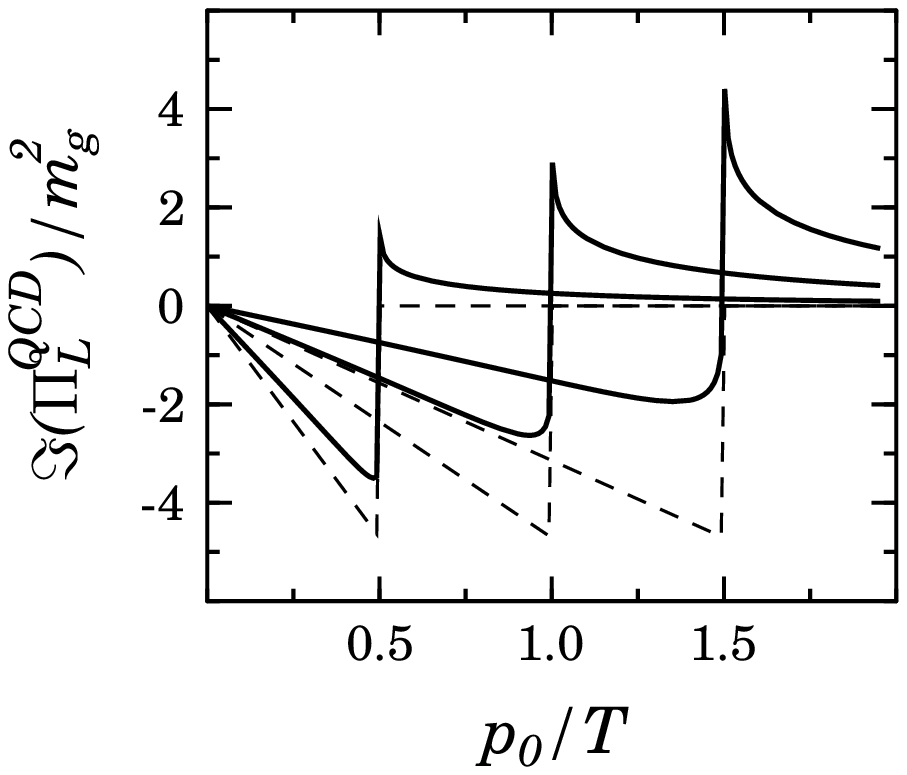}
 \caption{Imaginary part of the longitudinal gluon self energy within the complete one-loop
 approximation in Feynman gauge (solid line) and the HTL limit (dashed line) for
 $p/T=0.5, 1, 1.5$.}
\end{figure}

\begin{figure}
 \epsfxsize=10cm
 \epsffile{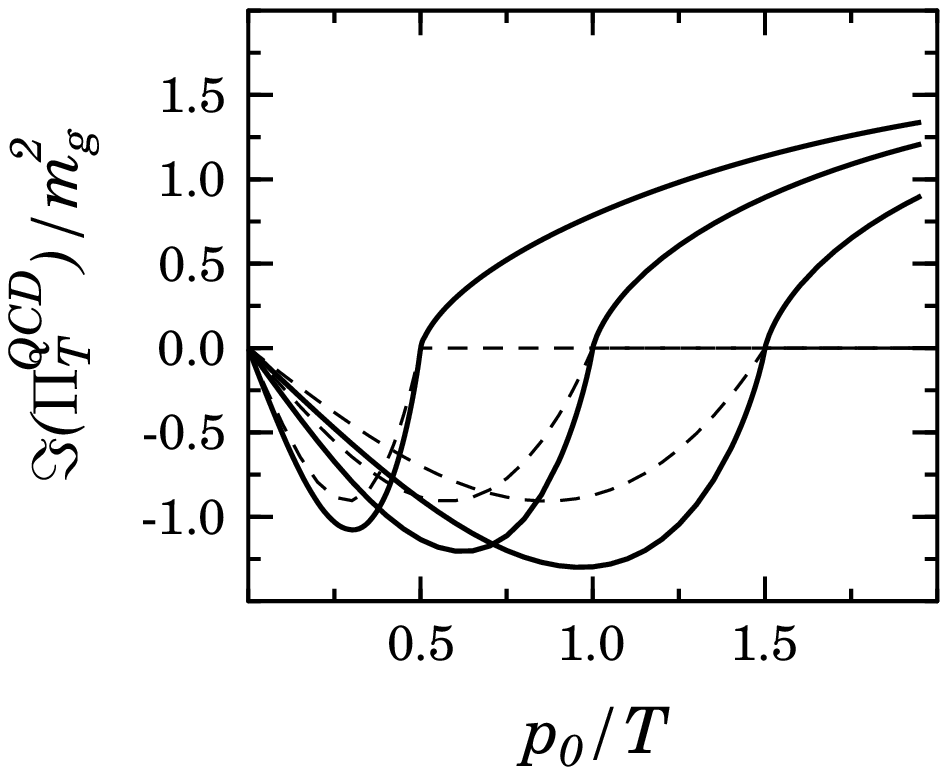}
 \caption{Imaginary part of the transverse gluon self energy within the complete one-loop
 approximation in Feynman gauge (solid line) and the HTL limit (dashed line) for
 $p/T=0.5, 1, 1.5$.}
\end{figure}

\begin{figure}
 \epsfxsize=10cm
 \epsffile{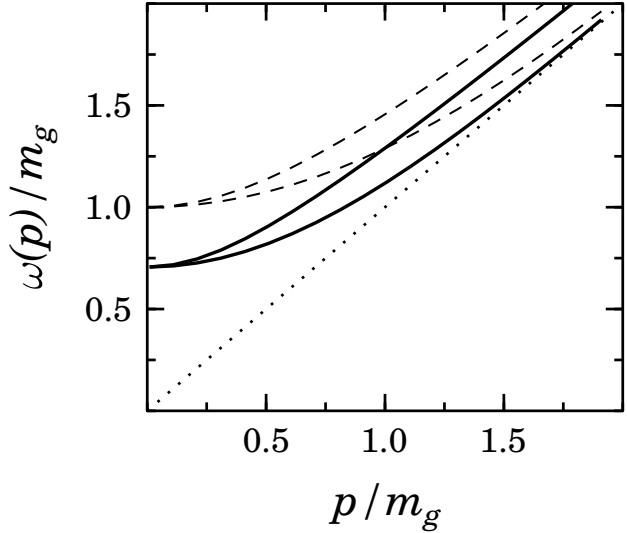}
 \caption{Dispersion relation of the longitudinal and transverse gluon modes
 (lower and upper branches, respectively) in one-loop approximation in Feynman gauge
 for $g=3$ (solid line) and HTL limit (dashed).}
\end{figure}

\begin{figure}
 \epsfxsize=10cm
 \epsffile{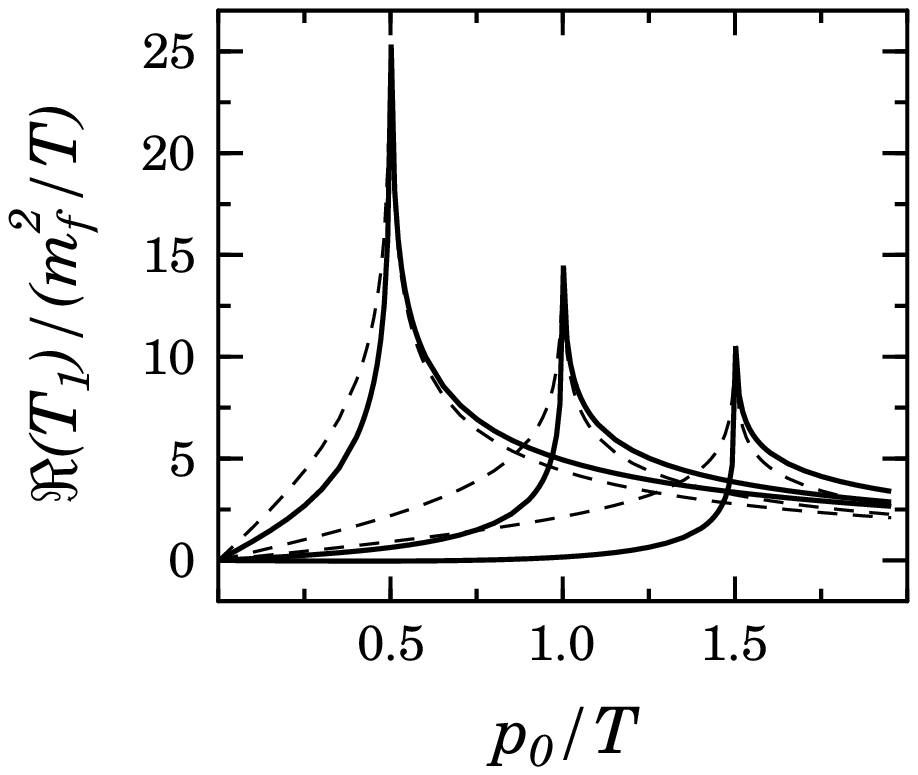}
 \caption{Real part of $T_1$ within the complete one-loop approximation in
 Feynman gauge (solid line) and the HTL limit (dashed line) for $p/T=0.5, 1, 1.5$.}
\end{figure}

\begin{figure}
 \epsfxsize=10cm
 \epsffile{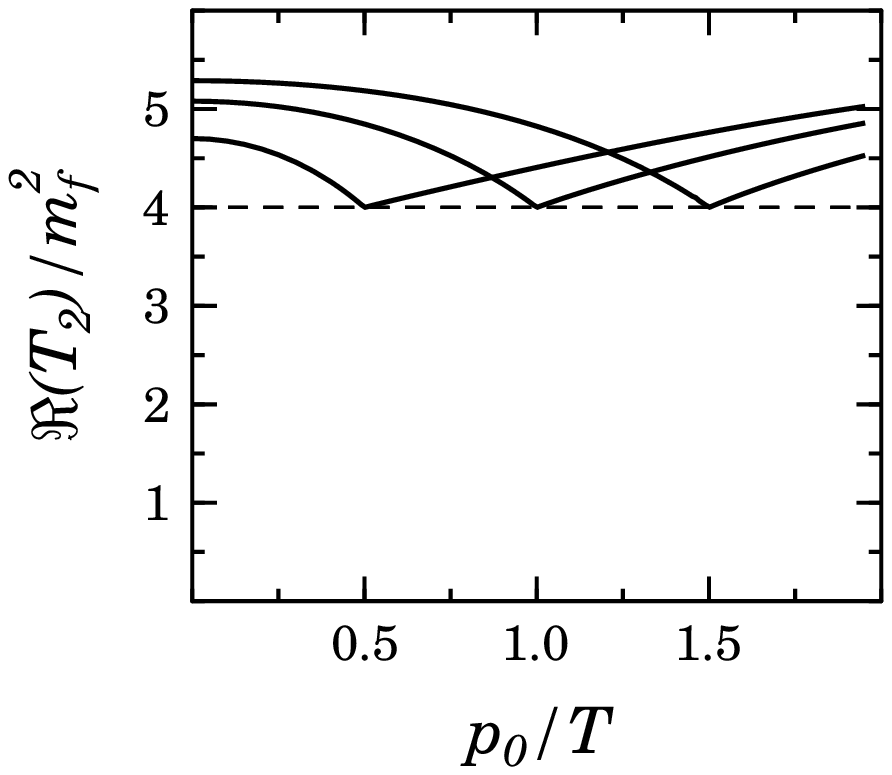}
 \caption{Real part of $T_2$ within the complete one-loop approximation in
 Feynman gauge (solid line) and the HTL limit (dashed line) for $p/T=0.5, 1, 1.5$.}
\end{figure}

\begin{figure}
 \epsfxsize=10cm
 \epsffile{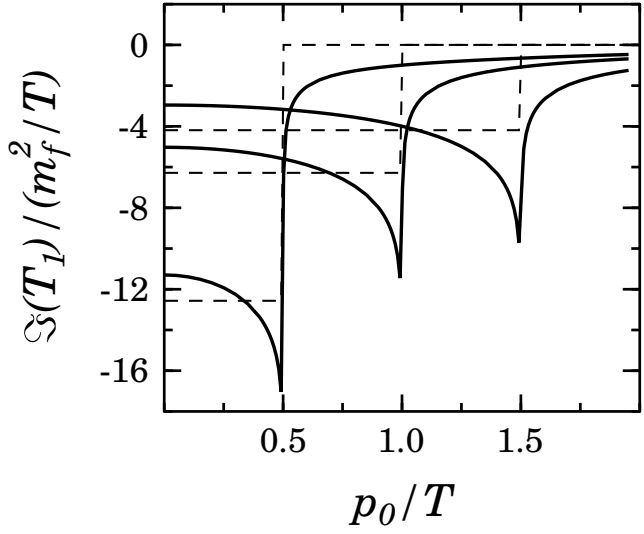}
 \caption{Imaginary part of $T_1$ within the complete one-loop approximation in
 Feynman gauge (solid line) compared to HTL result (dashed line) for $p/T=0.5, 1, 1.5$.}
\end{figure}

\begin{figure}
 \epsfxsize=10cm
 \epsffile{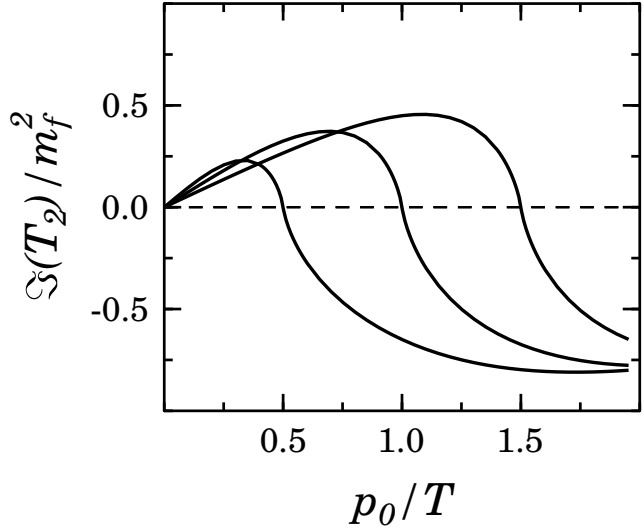}
 \caption{Imaginary part of $T_2$ within the complete one-loop approximation
 in Feynman gauge (solid line) for $p/T=0.5, 1, 1.5$. $T_2^{\rm HTL}$ has no imaginary part.}
\end{figure}

\begin{figure}
 \epsfxsize=10cm
 \epsffile{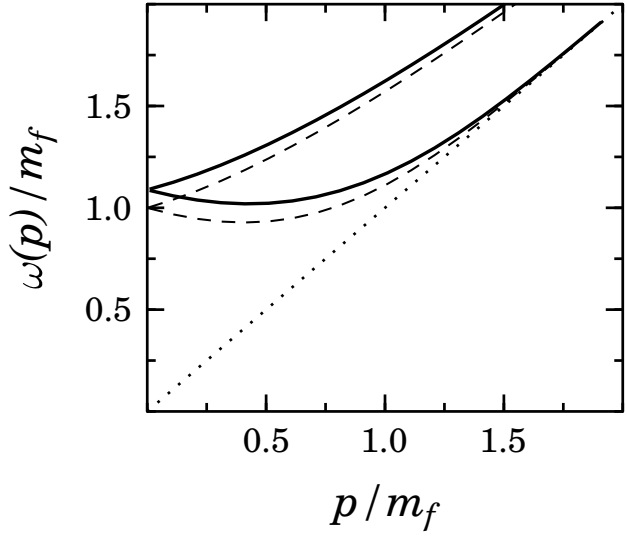}
 \caption{Quark dispersion relations within the complete one-loop
 approximation in Feynman gauge for $g=3$ (solid line) and the HTL limit (dashed line).
 The upper branch is the particle exitation, the lower one is the plasmino.}
\end{figure}

\begin{figure}
 \epsfxsize=10cm
 \epsffile{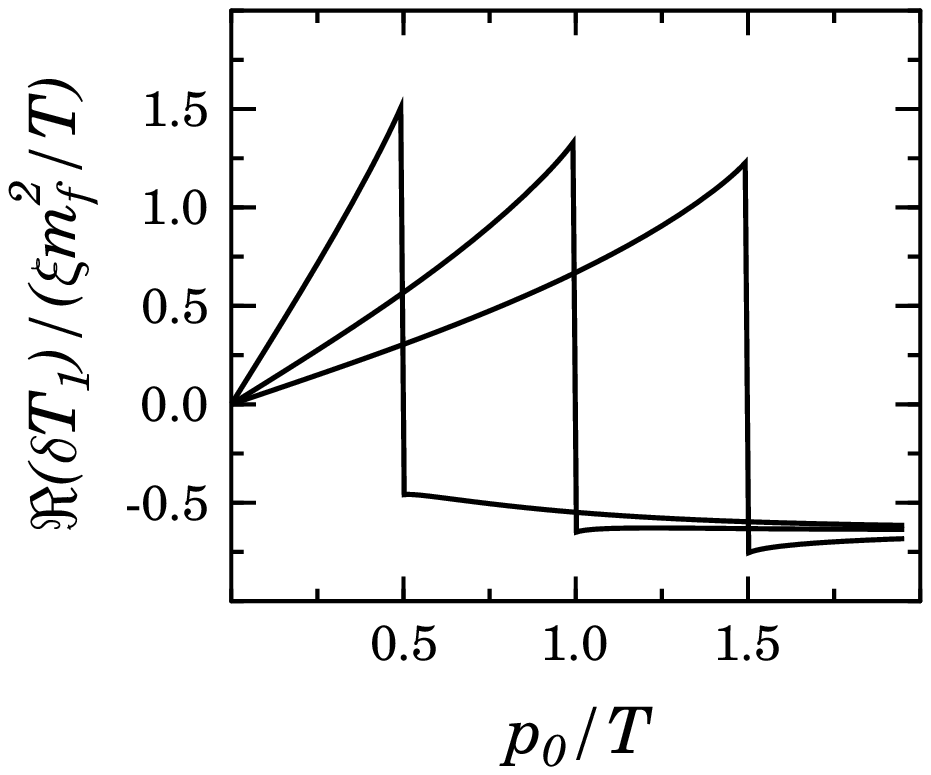}
 \caption{Gauge dependent contribution of the the real part of $T_1$ ($p/T=0.5, 1, 1.5$).}
\end{figure}

\begin{figure}
 \epsfxsize=10cm
 \epsffile{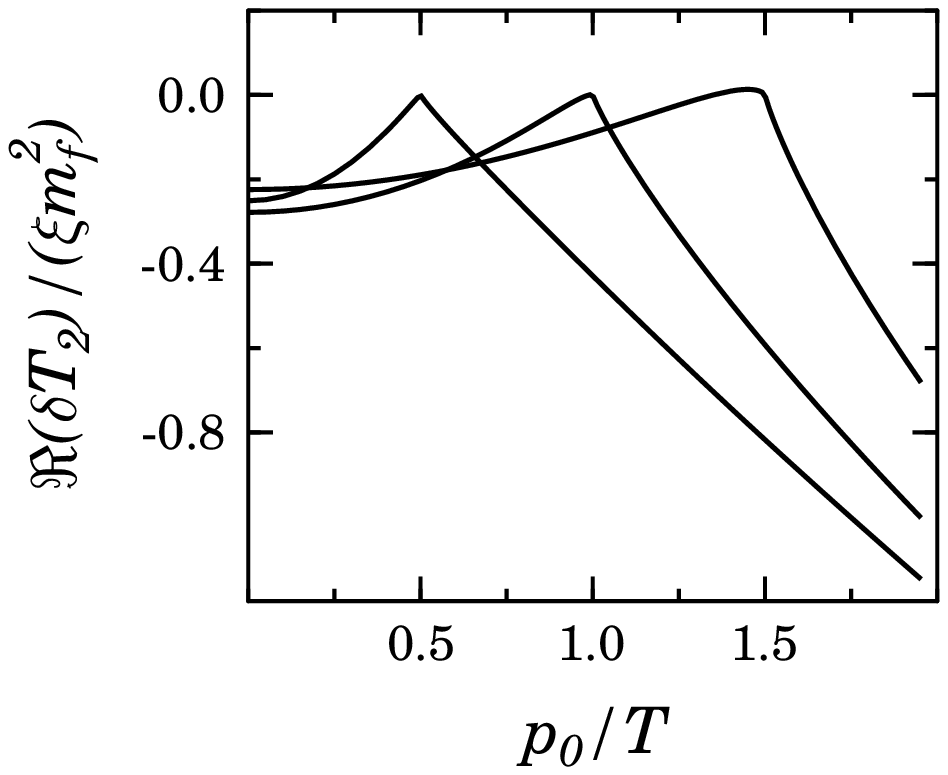}
 \caption{Gauge dependent contribution of the the real part of $T_2$ ($p/T=0.5, 1, 1.5$).}
\end{figure}

\end{document}